\documentclass{iopart}

\usepackage{epsfig}

\newcommand{\doublefigure}[2]{
  \begin{center}
    \begin{tabular}{cc}
      \epsfig{file=#1, width=0.465\textwidth} &
      \epsfig{file=#2, width=0.465\textwidth} \\
      \mbox{\sl(a)} & \mbox{\sl(b)} 
    \end{tabular}
  \end{center}
  }

\newcommand{\threefigure}[3]{
  \begin{center}
    \begin{tabular}{cc}
      \epsfig{file=#1, width=0.465\textwidth} &
      \epsfig{file=#2, width=0.465\textwidth} \\
      \mbox{\sl(a)} & \mbox{\sl(b)} 
    \end{tabular}
    \epsfig{file=#3, width=0.465\textwidth} \\
    \mbox{\sl(c)}
  \end{center}
  }

\begin{document}

\jl{6}   % Submit to CQG
%\submitted
\newcommand{\eqn}[1]{(\ref{eqn:#1})}
\newcommand{\fig}[2]{\ref{fig:#1}{\sl #2}}

%\draft
\begin{flushright}  LAUR-98-5276  \end{flushright}
\title[]{Apparent horizons in simplicial Brill wave initial data}

\author{Adrian P. Gentle\footnote{Permanent Address: Department of
  Mathematics and Statistics, Monash University, Clayton, Victoria
  3168, Australia. Email address adrian@lanl.gov.},
  Daniel E. Holz and Warner A. Miller}
\address{Theoretical Division (T-6, MS B288),\\ Los Alamos National 
  Laboratory, Los Alamos, NM 87545, USA}
\author{John A. Wheeler}
\address{Physics Department,\\ Princeton University, Princeton, NJ 08544, USA}

\begin{abstract}
We construct initial data for a particular class of Brill wave metrics
using Regge calculus, and compare the results to a corresponding
continuum solution, finding excellent agreement.  We then search for
trapped surfaces in both sets of initial data, and provide an
independent verification of the existence of an apparent horizon once
a critical gravitational wave amplitude is passed. Our estimate of
this critical value, using both the Regge and continuum solutions,
supports other recent findings.
\end{abstract}

%\maketitle
\setcounter{footnote}{0}

% --- BODY -----------------------------------------------------------

\section{Introduction}

Regge calculus \cite{regge61} is used to re-examine a particular class
of Brill wave spacetimes \cite{brill59} investigated by Miyama
\cite{miyama81}, and more recently, Holz \etal \cite{holz93} and
Alcubierre \etal \cite{alcubierre98}.  In particular, we study the
formation of apparent horizons within the simplicial initial data, and
confirm recent findings regarding the appearance of such horizons
\cite{alcubierre98}.

Brill wave initial data was first constructed using Regge calculus by
Dubal \cite{dubal89a}, although we show elsewhere \cite{gentle98b}
that his procedure is unable to capture the full structure of complex
axisymmetric initial data sets.  By using an improved lattice (which
is described in full in Ref. \cite{gentle98b}), we are able to
accurately reproduce the continuum solution, and find apparent
horizons in excellent agreement with recent two and three-dimensional
calculations \cite{alcubierre98}.

We proceed as follows.  In the next section, we briefly survey the
approach taken by Brill \cite{brill59} in the construction of
gravitational wave initial data.  In section \ref{sec:regge} we
describe the axisymmetric lattice used to construct Brill waves using
Regge calculus.  In section \ref{sec:trapped} we describe the method
used to find apparent horizons at a moment of time symmetry, and in
section \ref{sec:horizons} we investigate the appearance of apparent
horizons in the Brill wave space as the wave amplitude is increased.

\section{The continuum model}
\label{sec:continuum}

\begin{table}
\begin{center}
\begin{tabular}{|c|c|c|c|c|}     \hline
  a & $M_R$ & $M_C$  & $M$ (Holz \etal) &  $M$ (Alcubierre \etal) \\ 
  \hline 
  1  & $3.38 \times 10^{-2}$ & $3.38 \times 10^{-2}$ 
  & $3.40 \times 10^{-2}$ & $3.38 \times 10^{-2}$ \\
  2  & $1.26 \times 10^{-1}$ & $1.26 \times 10^{-1}$ & $1.26\times 10^{-1}$ 
  & $1.27 \times 10^{-1}$ \\
  5  & $6.98\times10^{-1}$ & $6.98\times 10^{-1}$ & $6.96 \times 10^{-1}$ 
  & $7.00 \times 10^{-1}$ \\
  10 & 2.91 & 2.91 & $2.91$ & $2.91$                \\
  12 & 4.67 & 4.67 & $4.67$ & $4.68$                \\ 
  \hline
\end{tabular}
\end{center}
\caption{Mass estimates for the Regge ($M_R$) and  continuum solutions
  ($M_C$).  The mass is calculated from the decay of $\psi$ in the
  asymptotic region.  We also show the results of previous
  calculations by Holz \etal \cite{holz93} and Alcubierre \etal
  \cite{alcubierre98}, which are in excellent agreement. All our
  results were calculated on a $600\times600$ grid.}
\label{tab:mass}
\end{table}

We examine the class of axisymmetric vacuum spacetimes containing
gravitational waves which admit a moment of time symmetry originally
studied by Brill \cite{brill59}.  Brill \cite{brill59} relaxed the
conformal flatness criteria and introduced a spatial metric of the
form
\begin{equation}
  \label{eqn:brill}
  ds^2 = \psi^4 \left\{ e^{2q} ( d\rho ^2 +dz^2 ) +
    \rho^2 \, d\phi^2 \right\},
\end{equation}
where the arbitrary function $q(\rho,z)$ can be considered the
distribution of gravitational wave amplitude \cite{wheeler64}.  In
order to obtain an axisymmetric solution which is reflection symmetric
about $z=0$, we require $q_{,\rho} = 0$ along $z=0$ and $q = q_{,z} =
0$ on $\rho=0$, together with the condition that $q$ falls off
sufficiently quickly (at least $r^{-2}$) in the asymptotic zone. Apart
from these boundary conditions, $q(\rho,z)$ is arbitrary.

At a moment of time symmetry the constraint equations of General
Relativity reduce to the single condition
\begin{equation}  \label{eqn:continuum_ivp}
  R = 0,
\end{equation}
where $R$ is the three-dimensional Ricci scalar.  With Brill's choice
of background metric, the Hamiltonian constraint
\eqn{continuum_ivp} takes the form
\begin{equation}  \label{eqn:brill_hamiltonian}
  \bigtriangledown ^2 \psi = - \frac {\psi}{4} \left( \frac {\partial
      ^2 q}{\partial \rho ^2} + \frac {\partial ^2 q}{\partial z ^2}
  \right)
\end{equation}
which is solved for $\psi (\rho,z)$ once $q(\rho,z)$ is given.  In the
current work we adopt the form of $q(\rho,z)$ chosen by Miyama
\cite{miyama81},
\begin{equation}   \label{eqn:q}
  q = a \, \rho ^2 e^{-r^2} \qquad 
\end{equation}
where $r^2 = \rho^2 + z^2$, and the wave amplitude $a$ is a free
parameter.

Since we require an axisymmetric, asymptotically flat solution which is
reflection symmetric on the $z=0$ axis, we enforce the  boundary
conditions 
\begin{equation}  \label{eqn:brill_boundary1}
  \frac {\partial \psi}{\partial \rho}(0,z) = 0, \qquad \quad
  \frac {\partial \psi}{\partial z}(\rho,0) = 0,  
\end{equation}
on $\psi$, together with a Robin condition at the outer boundary,
\begin{equation}
  \label{eqn:brill_boundary2}
  \frac {\partial \psi}{\partial r} = \frac{ 1-\psi
    } {r},
\end{equation}
which ensures that the hypersurface is asymptotically Euclidean.

\begin{figure} 
  \doublefigure{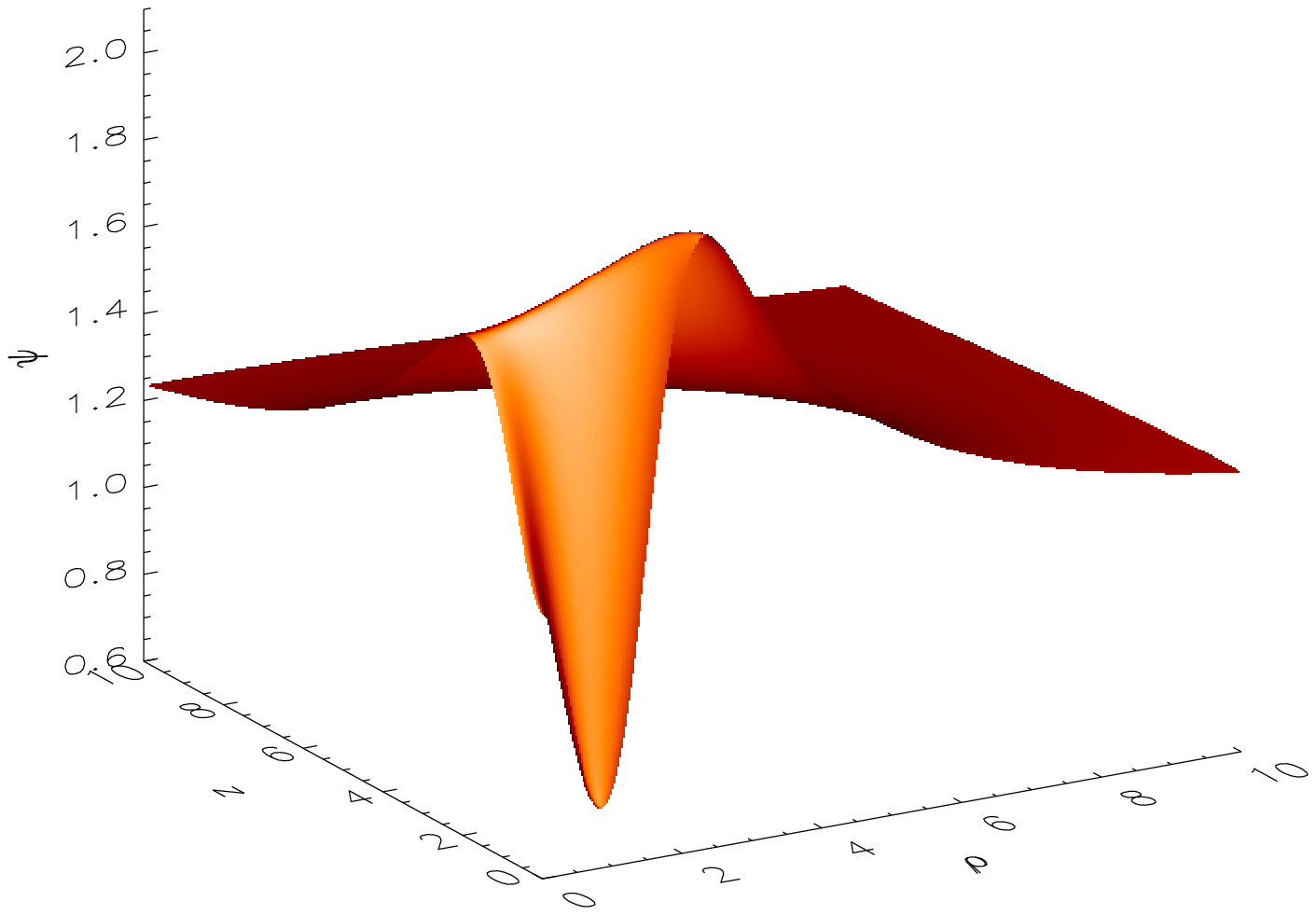}{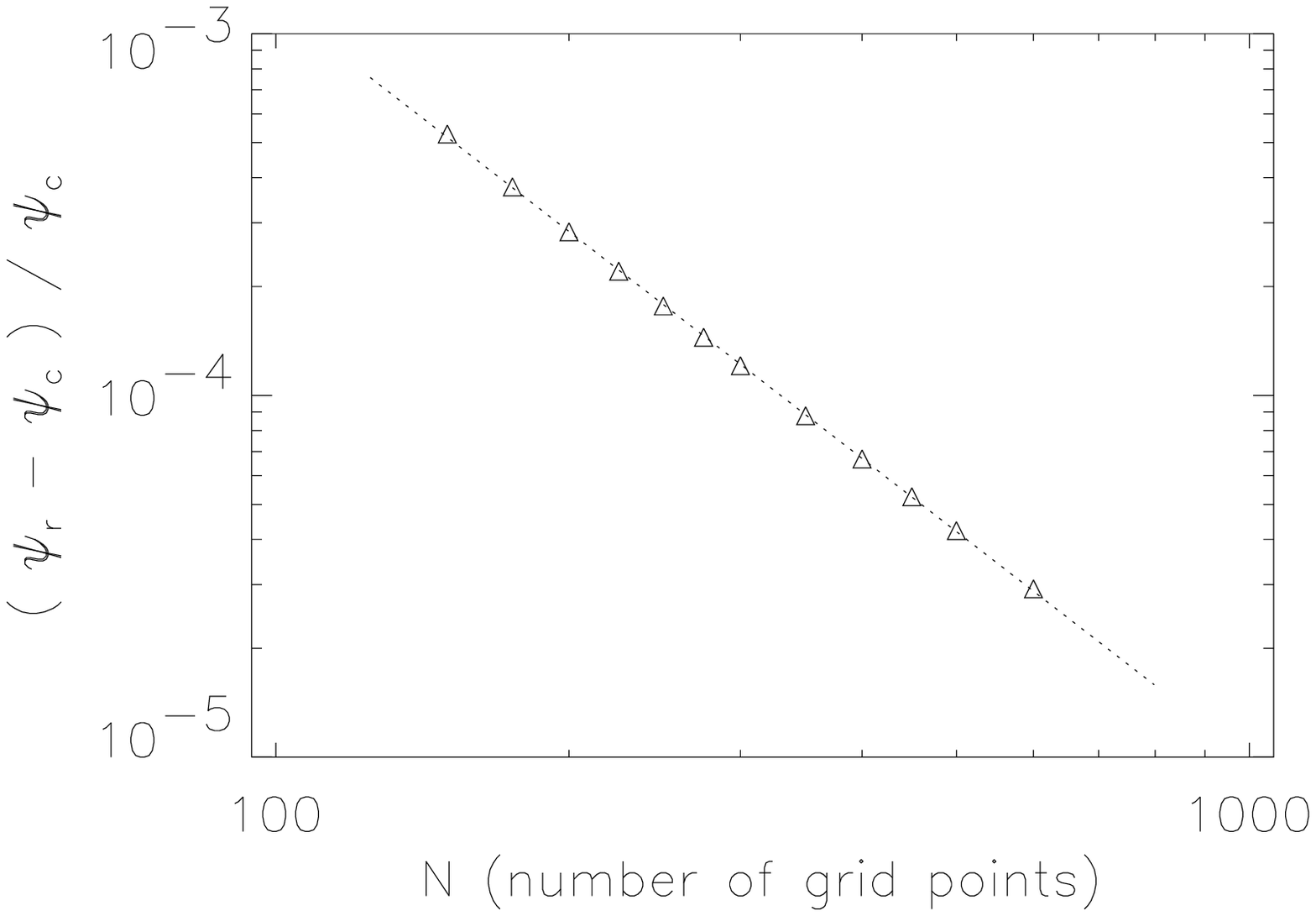}
  \caption{{\sl(a)} The conformal factor for $a=12$ calculated using
    the Regge lattice with $600\times600$ vertices.  The solution is
    shown on the first quadrant of the $\rho-z$ plane, with the outer
    boundaries placed at $\rho = z = 20$.  {\sl(b)} The fractional
    difference, averaged pointwise across the domain, between the
    simplicial ($\psi_r$) and continuum ($\psi_c$) solutions for
    the conformal factor on an $N\times N$ grid.}
  \label{fig:psi}
\end{figure}

The solution for $\psi$ is calculated using these boundary conditions
together with a second order accurate finite difference approximation
to equation \eqn{brill_hamiltonian}.

\section{The Regge calculus model}
\label{sec:regge}

In this section we will briefly outline the axisymmetric tetrahedral
lattice used to construct Regge initial data.  Full details of the
approach described here may be found elsewhere \cite{gentle98b}.

The simplicial equivalent of the initial value constraint at a moment
of time symmetry, equation \eqn{continuum_ivp}, has been given by
Wheeler \cite{wheeler64},
\begin{equation}  \label{eqn:regge_ive}
\sum_b L_{ab} \, \epsilon _{ab} = 0.
\end{equation}
This is the three-dimensional simplicial equivalent of $R=0$ at a
vertex $a$, where $L_{ab}$ is the lattice edge joining vertex $a$ to
vertex $b$, and $\epsilon_{ab}$ is the deficit angle (product of the
curvature on the edge by the area dual to the edge) about
$L_{ab}$. The summation is over all edges $L_{ab}$ which meet at
vertex $a$.

We mirror the continuum approach and perform a conformal decomposition
on the lattice edges, since they are the simplicial equivalent
of the metric.  This gives
\begin{displaymath}
  L_{ab} = \psi^2_{ab} \, \bar L_{ab}
\end{displaymath}
where $\bar L_{ab}$ is a freely chosen base edge length.  It is most
natural to define the simplicial conformal factor on the vertices of
the lattice, such that $\psi_a$ is the conformal factor at vertex $a$.
The conformal factor $\psi_{ab}$ acting on the edge $L_{ab}$ is then
defined using a centred, second order approximation, 
\begin{equation}
  \psi _{ab} = \frac{1}{2} \left( \psi_a + \psi_b \right).
\end{equation}

\begin{figure}
  \begin{center}
    \epsfig{file=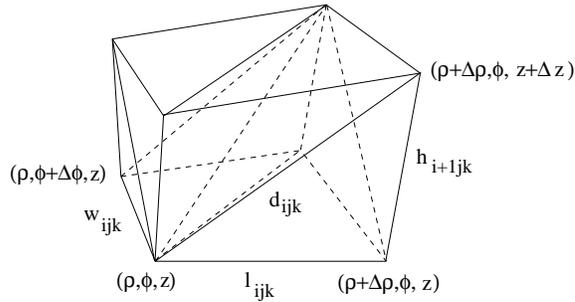,width=3.0in}
  \end{center}
  \caption{A portion of the simplicial three-lattice, which shows how
    each co-ordinate block is subdivided into six tetrahedra.  After
    calculating the geometric quantities for this lattice, the limit
    $\Delta \phi \rightarrow 0$ is taken, in order to obtain a
    discretization suitable for constructing axisymmetric initial
    data.}
  \label{fig:lattice}
\end{figure}

We assign the base edges $\bar L_{ab}$ such that the lattice is
aligned with the co-ordinate grids of the continuum approach.  Figure
\fig{lattice}{} indicates how we subdivide each block in the
co-ordinate grid to obtain a tetrahedral three geometry.  Given the
full set of physical edge lengths in this lattice, we calculate the
deficit angles (curvatures) about each edge, and write down the
simplicial initial value constraint, equation \eqn{regge_ive}.  This
equation is then solved, together with suitable boundary conditions,
for the conformal factor at each vertex of the lattice.

To obtain a purely axisymmetric lattice, we take the limit as $\Delta
\phi$ tends to zero, while demanding that each block is symmetric
about $\phi+\frac{1}{2}\Delta\phi$.  Retaining only the leading order
terms of equation \eqn{regge_ive} in this limit yields the axisymmetric
simplicial initial value equation at a moment of time symmetry.  Full
details of this calculation and the resulting equations may be found
elsewhere \cite{gentle98b}.  

By aligning the lattice with a cylindrical polar co-ordinate system,
we may use the form of the continuum base metric, equation
\eqn{brill}, to specify the base edges in the lattice.  Combining this
with the conformal decomposition described above, and noting that
after the limit has been taken there are only three non-zero edges per
vertex, the physical lattice edges take the form
\begin{eqnarray}  \label{eqn:conformal_cylindrical}
  l_{ik} & = & \left[ \psi ^2 \, e^q \right] _{i+\frac{1}{2},k} \,
  \Delta \rho _i   \nonumber \\
  d_{ik} &=& \left[ \psi ^2 \, e^q \right] _{i+\frac{1}{2}k+\frac{1}{2}}
  \left(\Delta \rho _i ^2  + \Delta z_k^2\right)^{1/2} \\
  h_{ik} & = & \left[ \psi ^2 \, e^q \right] _{i,k+\frac{1}{2}} \,
  \Delta z_k \nonumber 
\end{eqnarray}
where all expressions are centred on the edge, and we have neglected the
redundant index along the $\phi$ axis.

We are now able to calculate the simplicial initial data by solving
equation \eqn{regge_ive} for the conformal factor $\psi$.  We use the
same boundary conditions as before, applied along the $\rho=0$ and
$z=0$ boundaries using power series expansions into the domain.

The solution for $\psi$ calculated using the Regge lattice, together
with the boundary conditions described in section \ref{sec:continuum},
is shows in figure \fig{psi}{a}.  The solution displays excellent
agreement with the previous calculations of Holz \etal and Alcubierre
\etal \cite{holz93,alcubierre98}, and the continuum model developed in
section \ref{sec:continuum}.  Figure \fig{psi}{b} contains a rigorous
convergence analysis which demonstrates that the Regge solution
converges to the numerical continuum solution as the second power of
the lattice spacing.  For further details, we refer the reader to
Gentle \cite{gentle98b}.

Both solutions display good agreement with the previous calculations
of both Holz \etal \cite{holz93} and Alcubierre \etal
\cite{alcubierre98}.  Table \ref{tab:mass} contains estimates of the
mass of the Brill waves, calculated using a least squares fit of the
function
\begin{equation}
  \psi_a = 1+ \frac{M}{2r}
\end{equation}
to $\psi$ in the asymptotic zone.  The mass $M$ is found to converge
rapidly with radius.  The table also contains previous mass estimates
\cite{holz93,alcubierre98}, which again agree remarkably well with the
current calculations.

\section{Trapped surfaces}
\label{sec:trapped}

Our goal in the remainder of this paper is to locate any apparent
horizon which exists in the Brill wave data.  Apparent horizons are
defined \cite{hawking} as the outermost trapped surface on the spatial
hypersurface. That is, the outermost closed, spacelike two-surface on
which the divergence of outgoing null rays is zero.  At a moment of
time symmetry, this is equivalent to the outermost closed spacelike
surface with extremal surface area.

In order to construct the trapped surfaces we utilize the technique
employed by Holz \etal \cite{holz93}, where the search for an extremal
two-surface was shown to be equivalent to the construction of
geodesics in a certain two-dimensional space.

Let us consider a path $s(\rho,z)$ in the first quadrant of the
$\rho-z$ plane, which forms a closed surface when rotated about the
$z$ and $\rho$ axes.  The area of this surface of rotation is
\begin{equation}
  S = \int_{P_0}^{P_1} 2 \pi \psi^4 e^q \rho \left( d\rho^2 + dz^2
  \right) ^{1/2},
\end{equation}
and the criteria for a trapped surface is that this area is extremal,
$\delta S = 0$.  It is straightforward to show that this extremization
is mathematically equivalent to finding the geodesics of the metric
\cite{holz93}
\begin{equation} \label{eqn:false_metric}
  ds^2 = 4\pi^2 \psi^8 \rho^2 e^{2q} \left( d\rho^2 +dz^2 \right), 
\end{equation}
subject to appropriate boundary conditions. If the geodesics are
parameterized by $\lambda$, the appropriate boundary conditions are 
\begin{eqnarray}
  \dot\rho (\lambda_0) & = 0 \quad \mbox{and} \quad z(\lambda_0)=0,\\
  \dot z(\lambda_1) & = 0 \quad \mbox{and} \quad \rho(\lambda_1)=0,
\end{eqnarray}
where $\lambda_0$ and $\lambda_1$ are the endpoints of the curve, and
$\dot \rho$ represents differentiation with respect to $\lambda$.
These boundary conditions ensure that the curves are continuously 
differentiable across the axes.

The trapped surfaces are constructed using a shooting method. Given a
value of $\rho(\lambda_0)$, the geodesic is traced to $\lambda_1$,
where we check how closely the boundary condition is satisfied.  Once
a pair of  $\rho(\lambda_0)$ values have been found which bound the
condition $\dot z(\lambda_1)=0$, the bisection method  is used to
converge on the trapped surface.

It would be interesting to construct an algorithm for finding trapped
surfaces entirely within the framework of Regge calculus. An obvious
approach at a moment of time symmetry would be to implement a direct
search algorithm which minimizes the simplicial surface area of a
topological two-sphere.  For the purposes of the current work, we have
assumed that the simplicial lattice corresponds to a section of a
differentiable manifold, upon which we can apply the ideas outlined
above.  We use the conformal factor $\psi_r$, calculated using Regge
calculus, to construct the geodesics of metric \eqn{false_metric}, and
hence construct trapped surfaces within the lattice.

\section{Apparent horizons}
\label{sec:horizons}

\begin{figure}
  \threefigure{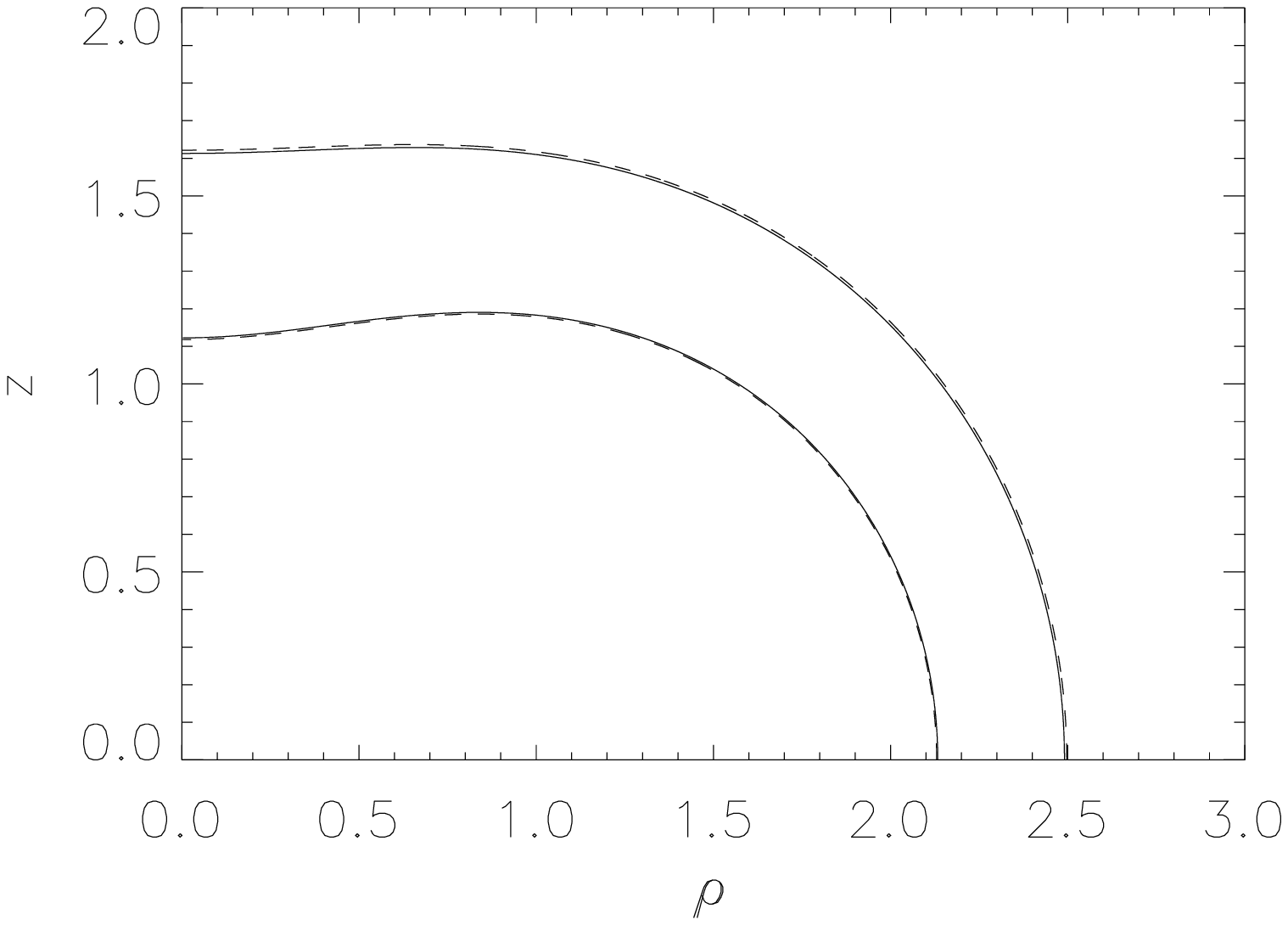}{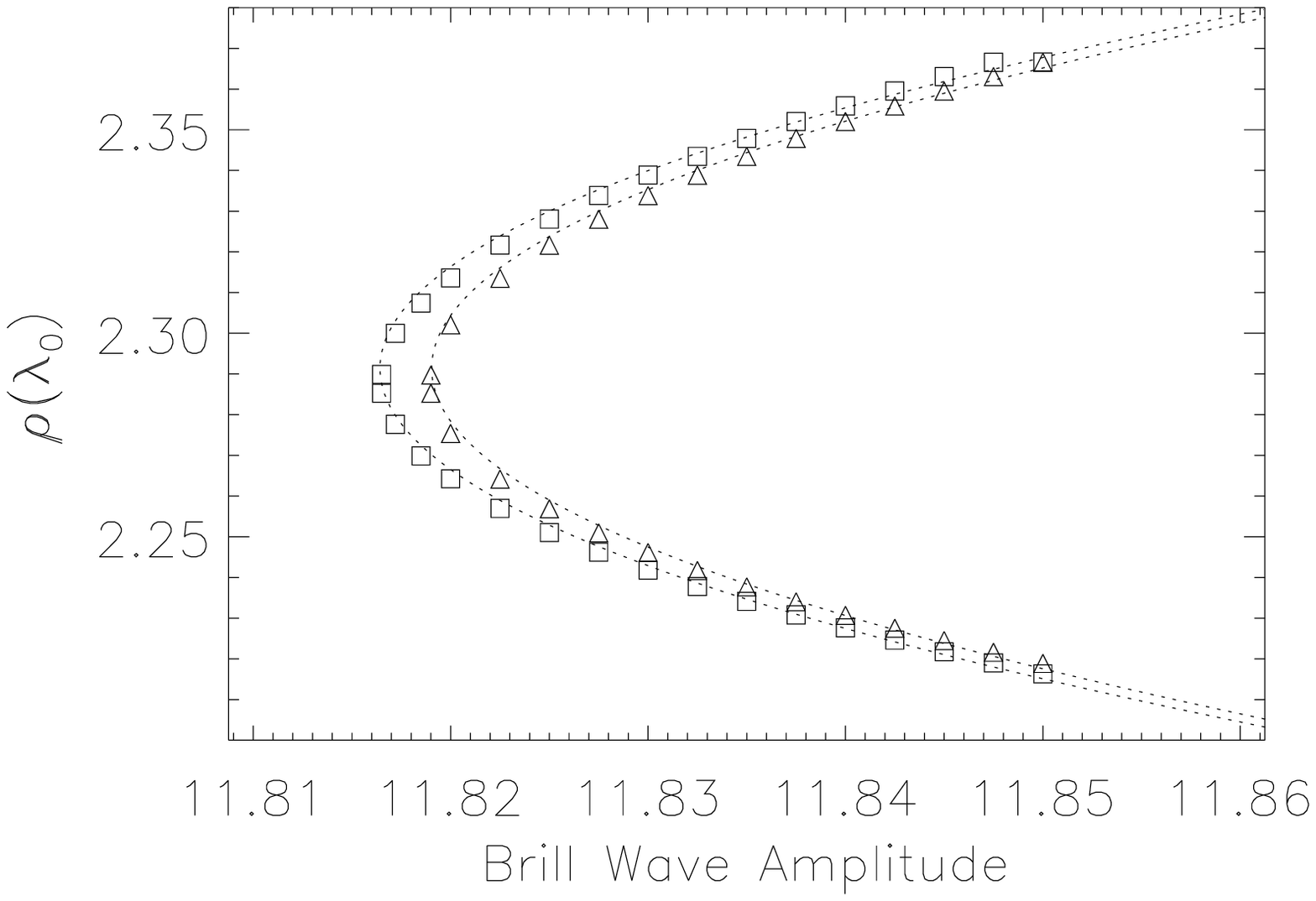}{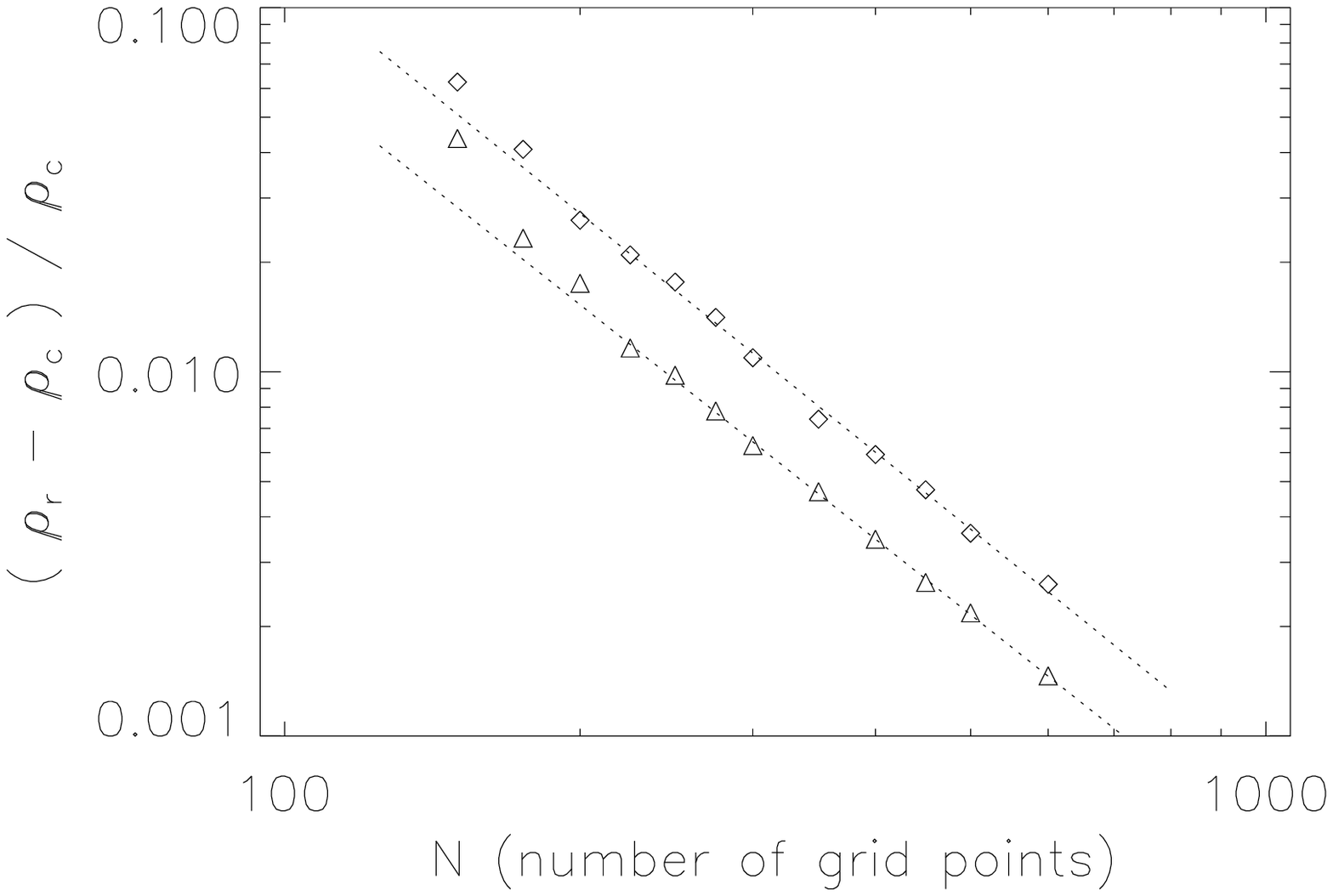}
  \caption{{\sl(a)} Trapped surfaces for a Brill wave with amplitude
    $a=12$ at a moment of time symmetry. The outermost trapped surface
    is the apparent horizon. The calculations were performed using
    both the Regge (solid lines) and continuum (dashed lines)
    solutions on a $600\times600$ grid, and are in excellent
    agreement. {\sl(b)} The point $\rho (\lambda_0)$ where the inner
    and outer trapped surfaces meet the $\rho$-axis is plotted as a
    function of wave amplitude, close to the critical amplitude $a_*$.
    The quadratic fit suggests that $a_* \approx 11.819$ in the Regge
    solution (triangles), and $a_*\approx 11.816$ in the continuum
    initial data (squares).  {\sl(c)} Convergence of the trapped
    surfaces calculated using the two sets of initial data.  For a
    Brill wave amplitude of $a=12$, we show the fractional difference
    between the Regge and continuum values of $\rho(\lambda_0)$ as the
    grid resolution is increased. Both the inner and outer trapped
    surfaces are shown.  The two solutions converge as the second
    power of the grid spacing. }
  \label{fig:horizons}
\end{figure}

Using the technique outlined in section \ref{sec:trapped}, we now
search for apparent horizons in the Brill wave initial data
constructed in sections \ref{sec:continuum} and \ref{sec:regge}.

We find that in both the Regge and continuum solutions an apparent
horizon first appears at a critical amplitude $a_*$, which lies in the
range $a_*\in [11.81,11.82]$.  As the amplitude $a$ is increased above
this critical value, two trapped surfaces are present, and these
gradually separate.  These results are in complete agreement with the
recent calculations of Alcubierre \etal \cite{alcubierre98}.

Figure \fig{horizons}{a} shows a projection of the trapped surfaces
found for a wave amplitude of $a=12$, calculated using both the
continuum and Regge solutions.  There is a slight difference between
the surfaces calculated using these two independent sets of initial
data, but the discrepancy is found to reduce as the second power of
the grid spacing.  In figure \fig{horizons}{b}, we show the position
$\rho(\lambda_0)$ of both trapped surfaces along the $\rho$-axis as a
function of the Brill wave amplitude.  The quadratic fit to this data
allows us to improve the estimate of the critical wave amplitude at
which an apparent horizon first forms.  We estimate that $a_* \approx
11.819$ for the simplicial solution, and $a_*\approx 11.816$ for the
solution to the continuum equation.  Both solutions predict that the
single horizon at $a=a_*$ passes through $\rho(\lambda_0) \approx
2.291$.

The value of the critical amplitude $a_*$, at which an apparent
horizon first forms, differs from that obtained in an earlier
calculation by some of us \cite{holz93}.  The discrepancy in the
earlier work was due to the relatively weak convergence criteria used
for the trapped surface finder.  Qualitatively, the previous work
agrees with both the present calculations and those of Alcubierre
\etal \cite{alcubierre98}.

\section{Future work}

We have used Regge calculus to confirm recent numerical findings
regarding the formation of apparent horizons in a particular class of
Brill wave spacetimes, while demonstrating that Regge calculus
provides an alternative and competitive technique for use in numerical
relativity. 

Brill wave spacetimes provide a challenging test-bed for the future
development of simplicial gravity.  Work is currently underway on the
development of a simplicial trapped surface finder for Regge calculus,
as well as the time evolution of the initial data constructed here, in
both $(2+1)$ and $(3+1)$-dimensions.  

\ack We gratefully acknowledge support from a Los Alamos National
Laboratory LDRD Grant.  One of us (APG) acknowledges support from the
Sir James McNeill Foundation at Monash University, and from the Center
for Nonlinear Studies.

% --- REFERENCES ----------------------------------------------------
\section*{References}

\gdef\journal#1, #2, #3, #4 {{#1}, {\bf #2}, #3 {(#4)}. }
\gdef\FP{\it{Found.~Phys.}}
\gdef\IJTP{\it{Int. J. Theor. Phys.}}
\gdef\IJMP{\it{Int. J. Mod. Phys.}}
\gdef\GRG{\it{Gen. Rel. Grav.}}
\gdef\PTP{\it{Prog. Theor. Phys.}}
\gdef\AP{\it{Ann. Phys.}}

\end{document}